\documentclass[a4,12pt]{article}
\begin{document}
\titlepage
\begin{flushright}
CERN-TH/2001-105\\
T/01-040\\
hep-th/0104077 \\
\end{flushright}
\vskip 1cm
\begin{center}
{ \Large
\bf Supersymmetric Flat Directions and Analytic Gauge Invariants}
\end{center}

\vskip 1cm
\begin{center}
{\large Ph. Brax\footnote{email: philippe.brax@cern.ch} }
\end{center}
\begin{center}
Theoretical Physics Division, CERN\\
CH-1211 Geneva 23\footnote { On leave of absence from  Service de Physique Th\'eorique, 
CEA-Saclay F-91191 Gif/Yvette Cedex, France}\\
\end{center}
\vskip .5 cm
\begin{center}
{\large C. A. Savoy \footnote{email: savoy@spht.saclay.cea.fr} }
\end{center}
\begin{center}
Service de Physique Th\'eorique, 
CEA-Saclay F-91191 Gif/Yvette Cedex, France\\
\end{center}

\vskip 2cm
\begin{center}
{\large \bf Abstract}
\end{center}
\vskip .3in \baselineskip10pt{We review some aspects of the
correspondence between analytic gauge invariants and supersymmetric
flat directions for vanishing D-terms and  propose a criterion to
include the F-term constraints.

}
\bigskip
\vskip 1 cm

\noindent
\newpage
\baselineskip=1.5\baselineskip
\section{\bf Flat Directions}

Flat directions in supersymmetric theories are continuously connected
degenerate supersymmetric vacua, modulo the group of gauge symmetry
$G,$ {\it i.e.,} each vacuum corresponds to a $G$-orbit. The degeneracy
of the classical solutions is described in terms of massless fields,
the so-called {\it moduli}.  The flat directions, {\it i.e.,} the moduli
space ${\cal V}$, define the low energy regime of the supersymmetric
theory. Their analysis has led to the conjecture of dualities [1]
between supersymmetric gauge theories as well as a new insight on 
confinement in some of such models [1,2,3].  Even in the presence of
low scale spontaneous supersymmetry breaking, the study of the flat
directions in the supersymmetric limit is still instrumental in many
applications.  The best example are the minimal supersymmetric
extensions of the Standard Model, where the flatness of the scalar
potential is lifted by the soft supersymmetry breaking operators.  In
general, the theory possesses other metastable vacua [4], which could
invalidate it -- or be important -- for phenomenological purposes.

The relevance of the analytic gauge invariants in the study of
supersymmetric theories was first  noticed a long time ago.  Indeed,
the $F$-terms in the scalar potential are components of the gradient of
the superpotential which is an invariant analytic function.  The
$D$-terms are Hermitean functions of the scalar fields, but a beautiful
result in algebraic geometry provides the link with holomorphic
invariants, at least for the analysis of the vacua [5,6,7,8].  The
fundamental mathematical property of the (classical) moduli space
(modulo the group of gauge symmetry) is its isomorphism with an
algebraic variety of analytic gauge invariant polynomials in the
primordial fields called the {\it chiral ring}.  They can be classified
in two categories, those which have algebraic constraints among the
invariants of the integrity basis, called {\it syzygies,} and those
with a free basis.  This classification is also related to the possible
patterns of gauge and global symmetry breaking.

Consider a set of chiral superfields $\phi^i$ whose scalar components
$z^i$ belong to a Kahler manifold with Kahler potential $K$. The scalar
potential of the globally supersymmetric theory with gauge group $G$ is
the sum $V= \frac{g_a^2}{2}D^{a2} + K^{i\bar j}F_i \bar F_{\bar j}$
where the $F$ terms are the gradients of the gauge invariant analytic
superpotential $W(z^i),$ {\it i.e.} $F_i=\partial_i W$, and the $D$
terms are $D^a= \partial_i K (T^a z)^i$ for the linear action $\delta
z^i=\epsilon_a (T^a z)^i$ of the compact gauge group $G$ on the scalars
$z^i$. The matrices $T^a$ are the generators of $G$.  The
supersymmetric vacua ${\cal V}$ of the theory are defined by the
conditions $F_i=0, \ D^a=0$. The gauge invariance of the Kahler
potential and the superpotential implies that these solutions form
G-orbits $\Omega=\{ gz_0,\ g\in G, z_0\in {\cal V}\}$.  The solution
$z$ has a little group $H_z$ in $G$. The $G$ invariance of the orbit
through $z$ implies that the little groups $H_{gz}=g^{-1}H_z g$ are
conjugate. The little groups are ordered according to the relation:
$H\le H'$ provided $H$ is conjugate to a subgroup of $H'$. Similarly
G-orbits are ordered according to their little groups. The orbits with
the largest little groups are called critical.

\subsection{\bf  D=0}
The $D$-flatness condition is non-analytic, the $F$-flatness condition
is analytic, i.e. involving only the scalars $z^i$ and not $\bar
z^{\bar i}$. The $D$-flatness conditions is non-trivially related to
analytic invariants.  Analytic invariants  $I(z)$  satisfy $\partial_i
I (T^{a}z)^i=0$. A necessary and sufficient condition for $D=0$ is the
existence of an analytic invariant $I$ and a non-zero constant $c$
such that

$$
\partial_i I=c\partial_i K. 
\eqno (1)
$$
These solutions are extrema of $I$ at fixed $K$. This correspondence
was pointed out in [6] and proved in [7], detailled discussions  can be
found in [8] and  [9].

\vskip 7pt
\noindent {\bf Example 1} -- Consider $G=SU(3)$ with three quarks
$Q_a^{\alpha}$ and three antiquarks $\bar Q_{\bar a}^{\bar\alpha}$,
$a=1\dots 3,\ \alpha=1\dots 3$. The most general invariant which is
linear in the basic invariants, see section 2, is  $ I=\rho_a
Q^{\alpha}_a \bar Q_{\bar a \alpha} +\mu
\epsilon_{\alpha\beta\gamma}\epsilon^{abc}Q^{\alpha}_a Q^{\beta}_b
Q^{\gamma}_c+ \tilde \mu \epsilon_{\bar \alpha\bar \beta\bar
\gamma}\epsilon^{\bar a\bar b\bar c}\bar Q^{\bar \alpha}_{\bar a} \bar
Q^{\bar \beta}_{\bar b} \bar Q^{\bar \gamma}_{\bar c}$ where $M_{a\bar
a}= Q^{\alpha}_a \bar Q_{\bar a \alpha}$, $B =
\epsilon_{\alpha\beta\gamma}\epsilon^{abc}Q^{\alpha}_a Q^{\beta}_b
Q^{\gamma}_c$ and $\bar B=\epsilon_{\bar \alpha\bar \beta\bar
\gamma}\epsilon^{\bar a\bar b\bar c}\bar Q^{\bar \alpha}_{\bar a} \bar
Q^{\bar \beta}_{\bar b} \bar Q^{\bar \gamma}_{\bar c}$.  General
solutions can be found from (1) and correspond to orbits of the
non-compact group $U(3,3)$[6]. They correspond to combinations of the
mesonic orbits:  $\partial_a  M_{a\bar a}=c ( Q^{\alpha}_{ a})^*$
yields $ (Q^1_1)^*=\bar Q_{11}$, and the baryonic orbits: $\partial_a
B=c ( Q^{\alpha}_{ a})^*$ yields $Q^1_1=Q^2_2=Q^3_3 ,$ all others zero
(and analogously for antibaryons), up to a $U(3,3)$ transformation. The
critical orbit $Q^1_1=Q^1_2=Q^1_3$, all others zero, breaks $ SU(3)\to
SU(2)$ and gives $I=0$ and $D_{U(1)}\ne 0$ for the broken  $U(1)$ in
$SU(2)\times U(1)\subset SU(3)$. It is not a flat direction and its
$U(3,3)$ orbit is not a D-flat direction too.

\vskip 7pt
\noindent {\bf Example 2} -- Consider $G=E_6$ with one representation
$27$. There is only one invariant $I=d_{abc}z^az^bz^c$. Two critical
orbits have little groups $SO(10)$ and $F_4$ respectively. For $z_0$ in
the $SO(10)$ critical orbit one finds $I(z_0)=\partial_i I(z_0)=0$ and
$D_{U(1)}(z_0,\bar z_0)\ne 0$ since $ z_0$ is charged under the $U(1)$
such that $SO(10)\times U(1) \in E_6$. For $z_0$ in the orbit with
little group $F_4$ one checks that $D=0$ with $I(z_0) \ne 0$. 

It is important to note that  critical orbits with a unique singlet
corresponding to a broken $U(1)$ such that $H_{z_0}\times U(1)\subset
G$ is maximal are not $D-$flat, as discussed below. These orbits will
be identified with the open orbits of the complexified gauge group
$G^c$.

\subsection{F=0}
The superpotential is $G$ invariant and since it is an analytic
function it is also   $G^c$ invariant where $G^c$ is the complexified
gauge group $G^c=\{ g=e^{i\alpha_a T^a},\ \alpha_a \in C\}$.  For each
solution of $F^i=0$ the whole complex orbit $\Omega^c=\{gz_0, g\in
G^c\}$ is a solution.  Note that if $z$ has a (complexified) little
group $H^c_z$, the number of non-trivial conditions $\partial_i W=0$ is
equal to the number of $H^c_z$ singlets.  Since $\partial_i W=0$, the
superpotential cannot be chosen to be the invariant $I$ leading to
solutions of $D=0$. For a generic superpotential $W(I^a)$ containing
all the basic invariants $I^a(z)$ with degree $d_a$, $F^i=0$ implies
that
$$
\sum_a  d_a
\frac{\partial W}{\partial I^a}I^a(z_0)=0.
\eqno(2)
$$
Now for $D=0$ there is at least one invariant such that $I(z_0)\ne 0$.
This implies that directions corresponding  to open critical orbits
with all $I^a(z_0)=0$ are $F$-flat  but not $D$-flat directions.  They
are characterized by a $H_{z_0}$ singlet in the coset $G/H_{z_0}$ such
that $H_{z_0}\times U(1)\subset G$. If $z_0$ is critical, it is the
only $H_{z_0}$ singlet, the action of the complexification  of the $U(1)$ is
$z_0 \rightarrow \lambda z_0 ,$  $\partial_{z_0}I^a$ is the only
potentially non-vanishing gradient direction, and $z_0\partial_{z_0}I^a
= z^i\partial_i I^a=d_a I^a$  vanishes.  Therefore $\partial_{z_0}I^a
=0$ for all invariants $I^a$. Now since (1) is necessary, $D=0$ can
only be satisfied for $z^0=0$ which is not in the open orbit but in its
closure. 

This provides counter-examples to the statement that for each
$G^c$-orbit which gives  solutions to $F^i=0$ one can always find one
($G$-orbit) solution of $D=0$.  For a generic $W$ and $z_0$ such that
$D=0$, the $F^i=0$ conditions  lift at least  one of the flat
directions.   If some of the $I^a$ are excluded from $W$, e.g. with
global symmetries, then they can be used to define $D$-flat directions
which are consistent with $F=0$.

\vskip 7pt
\noindent {\bf Example 3} -- Let us consider a simplified
supersymmetric version of the Standard Model, with gauge group
$SU(2)\times U(1),$ and the following chiral superfield content: 
two Higgs $SU(2)$ doublets $H_1 ,\, H_2 ,$ with $U(1)$ charges $Y= -1,
+1,$ resp., two lepton doublets $L_1 ,\, L_2 ,$ with $Y= -1,$ and two
lepton singlets  $E_1 ,\, E_2 ,$ with $Y= 2.$ We also introduce a
lepton parity, with $R=-1$ for leptons and $R=1$ for Higgses.  The most general
invariant (at most cubic) superpotential is of the form (up to some
redefinition), $W= \mu H_1  H_2 + \lambda_1 H_1  L_1 E_1 +
\lambda_1 H_1  L_2 E_2.$ The following invariants have been
excluded because of the $R-$parity: $L_1  H_2 ,$ $L_1  H_2 ,$
$E_1 L_1  L_2 ,$ and $E_2 L_1  L_2 .$ By taking a linear
combinations of these invariants and those in $W$ as the invariant $I$ in equation (1)
one finds  D-flat solutions. The moduli space is its intersection
with the solutions of $\partial_i W =0,$ and has components such that
$H_2$ and the lepton scalars are non-vanishing. This breaks completely
the gauge symmetry (in particular the electromagnetic $U(1)$), but also
the lepton parity. A similar situation is found when
families of quark multiplets are introduced with some kind of baryon or
lepton symmetry, giving rise to potentially dangerous charge and colour
breaking directions of the moduli space [4]. The flat directions of the
minimal supersymmetric extension of the Standard Model are listed in [5].

\section{Moduli and Syzygies}

To any solution of $D^a=0$ one can associate a holomorphic gauge
invariant satisfying (1).  The proof of (1) is obtained by studying the
closed orbits of the complexified $G^c$ of the gauge group $G$ and the
ring of $G^c$--invariant analytic polynomials.  This ring is finitely
generated: one can find an integrity basis, i.e.  a set of
$G$-invariant holomorphic homogeneous polynomials $\left\{ I^a (z)
\right\}_{a=1\cdots d}$ such that every $G^c$-invariant polynomial in
$z$ can be written as a polynomial in the $I^a (z)$.

Notice that orbits with all $I^a(z)=0$, $z\ne 0$, are open, i.e. there
exists a one parameter subgroup of $G^c$ with generator $T$ such that
$e^{\xi T}\in G^c$ and $e^{\xi T}z\to 0,\ \xi\to\infty$. The compact
generator associated to $T$, i.e. $iT$, corresponds to a broken $U(1)$
factor in the coset $G/H_z$ with maximal subgroup $H_z\times
U(1)\subset G$. In example 2, the $SO(10)$ orbit is open, $iT$ is the
$U(1)$ charge, while the $F_4$ orbit is closed.

The elements of  an integrity basis are not always algebraically
independent.  In general, there exist algebraic relations (called {\it
syzygies}) satisfied by the fundamental invariants ${\cal S}^{\alpha}
\left( I^a (z) \right) = 0$. In example 1, there is one such relation
$\det M- B {\tilde B}=0$.

To each closed $G^c$--orbit corresponds a vector in $C^d$ made out of
the values taken by the invariants $\left\{ I^a (z) \right\}$ along
this orbit and  satisfying the syzygies.  In that sense the algebraic
manifold defined by the syzygies\ is identified with the set of closed
$G^c$--orbits.  Notice that the origin $\left\{ I^a = 0 \right\}$ is
associated with the unique closed $G^c$--orbit of $z^i = 0$.

The existence of the syzygies can be related to the index of the matter
field representation,  $\mu$  defined by
$\hbox{tr}({T^aT^b})=\mu\delta^{ab}.$ For low indices $\mu<\mu_{adj},$
where  $\mu_{adj}$ is the index of the adjoint representation, it has
been shown that there are no syzygies.  For $\mu>\mu_{adj}$ the generic
situation is that there are syzygies, with a few exceptional cases with
no syzygies.

Equation (1) can be seen  as a condition for the points of a closed
$G^c$-orbit to extremize the K\"ahler potential, i.e. $\partial_i(\lambda_a I^a -K)=0$
with Lagrange multipliers $\lambda_a$.  A result of geometric
invariant theory [7] states that the points extremizing the
K\"ahler\ potential on a $G^c$-orbit form a unique $G$-orbit and are
solutions of $\left\{ D^A = 0 \right\}$.  Identifying   the points on a
same $G$-orbit,  there is a one-to-one correspondence between any two
of the following sets:  {\it (i) the algebraic manifold ${\cal M}_I$
defined by the syzygies; (ii) the closed $G^c$-orbits; (iii) the
solutions of (1) modulo gauge transformations; (iv) the solutions of
$D^A=0$ modulo gauge transformations.}

Now imposing the $F$-flatness condition implies  an extra
relation $\sum_a d_a \frac{\partial W}{\partial I^a}I^a=0$ to the
syzygies. Geometrically one intersects the hypersurface  deduced from
the superpotential with the moduli space ${\cal M}_I$. It is a
necessary condition. Three case are to be envisaged. If
there is no intersection the only solution is $z=0$, if it exists. If
$z=0$ is not a solution then supersymmetry is spontaneously broken. If
the intersection is reduced to one point the $D$-flat direction is
completely lifted.  Finally in the generic situation the intersection
has at most dimension $\hbox{dim}{\cal M}_I -1$ corresponding to the
lifting of at least one flat direction.

\section{\bf  Dualities and Confinement}

One of the most striking result on supersymmetric gauge theories is the
existence of a new type of duality. This duality relates two apparently
different theories in the short distance regime that are described by
the same effective theory in the infrared limit. In the same vein the
basic question concerning the  issue of colour confinement has been
tackled and clarified in these non-perturbative approaches for a large
class of supersymmetric theories.  Effective theories have been written
and argued to describe the IR behaviour of some gauge theories in terms
of gauge invariant composite chiral superfields [1,2,3].
The non-perturbative quantum effects
are fixed by the  holomorphy of the superpotential, by the global
symmetries of the theory and by several descent links among series of
such theories.

A powerful and necessary criterion for the existence of an effective
theory describing the IR regime of an asymptotically free gauge theory
was stated by 't~Hooft [10]:  there should be  matching of the (formal)
anomalies of the global symmetries calculated with either the UV or the
IR massless fermions. The analysis of the isomorphism between  the
moduli space and the chiral ring constrained by the syzygies leads to
the following (partially proved) conjecture [11,12] {\it
The 't~Hooft conditions are satisfied for supersymmetric gauge theories
if and only if the syzygies of the chiral ring of gauge invariants
derive from a superpotential} $W(I_a)$, {\it through} ${\cal
S}^a=\partial W/ \partial I^a =0$.

Now by comparing with the general structure of $W$ required by the
$R$-symmetries, one obtains the following condition:  {\it A necessary
condition for the matching of the anomalies is:} $\mu = \mu_{adj} + k,
\ (k=0,1,2).$ This is the confinement condition for theories with $\mu
\ge \mu_{adj}$ [13]. There are some subtleties and peculiarities that
can be found in the literature.

It is generally assumed that theories with $2+\mu_{adj}<\mu<3\mu_{adj}$
have an infrared fixed point [1] where they are described by a
superconformal theory. The syzygies are therefore exact quantum
relations between the chiral primary fields at the superconformal fixed
point.  Whereas in ``electric'' theories the gauge invariants are
composite fields and so subject to  syzygies, in the magnetic dual
theories, some of them appear as elementary fields and do not have
{\it{a priori}} to satisfy the syzygies. These ``magnetic syzygies''
appear as the equations of motion from the superpotential generated non
perturbatively in the magnetic theories [14].

These dualities amount to an identification between the moduli spaces,
{\it{i.e.}}, between the flat directions of the dual theories. It is
worth noticing that the D-flatness in a theory corresponds to the
F-flatness in the dual theory.

%%%%%%%%%%%%%%%%%%%%%%%%%%%%%%%%%%%%%%%%%%%%%%%%%%%%%%%%%%%%%%%%%%%%%%%%%%%%%%
%%%%%%%%%%%%%%%%%%%%%%%%bibliographie%%%%%%%%%%%%%%%%%%%%%%%%%%%%%%%%%%%%%%%%%%

\vskip 1.3cm
\noindent
[1]
N. Seiberg,
\newblock Nucl. Phys. B435 (1995) 129.

\noindent
[2]
T. R. Taylor, G. Veneziano and S. Yankelowiz, 
\newblock Nucl. Phys. B218 (1983) 493.

\noindent
[3]
N. Seiberg,
\newblock Phys. Rev. D49 (1994) 6857.

\noindent
[4] J. A Casas, A. Leyda and C. Mu\~{n}oz,
\newblock Nucl. Phys. B471 (1996) 3.

\noindent
[5]
T. Gherghetta, C. Kolda, S.P. Martin,
\newblock Nucl.Phys. B468 (1996) 37.

\noindent
[6]
F. Buccella, J.P. Derendinger, S. Ferrara and C.A. Savoy,
\newblock Phys. Lett. 115B (1982) 375.

\noindent
[7]
C. Procesi and G.W. Schwarz,
\newblock Phys. Lett. 161B (1985) 117.

\noindent
[8]
R. Gatto and G. Sartori,
\newblock Commun. Math. Phys. 109 (1987) 327.

\noindent
[9]
M.A. Luty and I. Washington~Taylor,
\newblock Phys. Rev. D53 (1996) 3399.

\noindent
[10]
G. 't~Hooft et~al., editors,
\newblock Naturalness, chiral symmetry breaking and spontaneous chiral
symmetry breaking, NATO advanced study, Cargese, France, Plenum, 1980.

\noindent
[11]
G. Dotti and A.V. Manohar,
\newblock Nucl. Phys. B518 (1998) 575.

\noindent
[12] Ph. Brax, C. Grojean and C. A. Savoy 
\newblock Nucl. Phys. B 561 (1999) 77.

\noindent
[13]
C. Cs{\'a}ki, M. Schmaltz and W. Skiba,
\newblock Phys. Rev. D55 (1997) 7840.

\noindent
[14] K. Intriligator and N. Seiberg,
\newblock Nucl. Phys. Suppl. 45BC (1996) 1.

\end{document}